 \documentclass[twoside,reqno,11pt]{fcaa}

\usepackage{graphicx}
\usepackage{epsfig}
\usepackage{amsthm}
\usepackage{amsmath}
\usepackage{latexsym}
\usepackage{amsfonts}
\usepackage{amssymb}

\newtheoremstyle{theorem}
  {15pt}          
  {15pt}  
  {\sl}  
  {\parindent}
  {\sc}  
  {. }    
  { }    
  {}     
\theoremstyle{theorem}

\newtheoremstyle{defi}
  {15pt}          
  {15pt}  
  {\rm}  
  {\parindent}     
  {\sc}  
  {. }    
  { }    
  {}     
\theoremstyle{defi}

 \def\proofname{\indent\rm P\ r\ o\ o\ f.}
 
 \def\qed{\hfill$\Box$} 

 \usepackage{hyperref} 

 \def\themycopyrightfootnote{\vspace*{3pt}
 \copyright \, Year\,  Diogenes  Co., Sofia
   \par  \noindent pp. xxx--xxx
   \hfill  \vspace*{-36pt}
  }

 \title[Generalized Erd$\acute{\text{E}}$lyi-Kober type integrals]{Towards  a geometric interpretation of generalized fractional integrals -  Erd$\acute{\text{E}}$lyi-Kober type integrals on $R^N$ as an example  \\ [4pt] }
 \author[R.Herrmann]{Richard Herrmann $^1$}

\begin{document}

 \vbox to 2.5cm { \vfill }


 \bigskip \medskip

 \begin{abstract}
A family of generalized Erd$\acute{\text{e}}$lyi-Kober type fractional integrals is interpreted geometrically as a distortion of the rotationally invariant integral kernel of the Riesz fractional integral in terms of generalized Cassini ovaloids on $R^N$. Based on this geometric view, several extensions are discussed.\\ 
 
\medskip

 \smallskip

{\it Key Words and Phrases}: Fractional calculus, Riesz fractional integrals, Erd$\acute{\text{e}}$lyi-Kober fractional integrals, Cassini ovaloids.

 \end{abstract}

 \maketitle

 \vspace*{-16pt}



\section{Introduction}
In the following we want to present a geometric approach for a deeper understanding of concepts and strategies used in generalized fractional calculus \cite{kir94}. 

We will collect arguments in support of  the idea, that a generalization of fractional calculus may be
considered from a geometrical point of view as a distortion of the isotropic kernel commonly used in standard fractional calculus, mediated by  one or more  additional fractional parameters.

E.g. a fractional integral $I^\alpha$ acting on a function $f(x)$ on $R^N$ is therefore generalized to a multi-parameter fractional integral, where the additional parameters are a measure of distortion:  
\begin{equation}
\label{cass1}
I^\alpha  f(\vec{x})   \rightarrow I^{\alpha, \gamma, ...}  f(\vec{x})
\end{equation}
According to \cite{go},  fractional  integrals are of convolution type and exhibit weakly singular kernels of power-law type. 

Therefore as a first step we will investigate in this paper the specific geometric properties of kernels or weight-functions of a generalized set of multi-dimensional fractional integrals of 
Erd$\acute{\text{e}}$lyi-Kober type. 

For this case, we will demonstrate,  that a geometric approach allows a direct classification and interpretation of generalized multi-parameter fractional integrals in a straight forward manner in terms of Cassini and Maxwell ovaloids. 

Furthermore, based on this viewpoint, we will present 
some generalizations of fractional operators of Erd$\acute{\text{e}}$lyi-Kober type, which allow a direct application in hadron physics.    

\section{Two examples as an illustration}
In one dimensional space we start with some examples to illustrate the procedure:
The Liouville definition of the left and right  fractional integral [\cite{lio32}] is given by:
\begin{eqnarray}
I_{+}^\alpha f(x) 
&=&  
\frac{1}{\Gamma(1-\alpha)}   
     \int_{-\infty}^x \!\!\!  d\xi \, (x-\xi)^{-\alpha} f(\xi)
 \\
I_{-}^\alpha f(x)
&=& 
\frac{1}{\Gamma(1-\alpha)}  
     \int_x^\infty  \!\!\!  d\xi \, (\xi-x)^{-\alpha} f(\xi)
\end{eqnarray} 
With the fractional  parameter  in the interval  $0 \leq \alpha \leq 1$. Consequently for the limiting case $\alpha=1$ $I_{+}$ and $I_{-}$ both coincide with the unit-operator and for $\alpha=0$  $I_{+}$ and $I_{-}$ both correspond to  the standard integral operator.

If $x$ is a time-like coordinate, the left Liouville integral is causal, the right Liouville integral is anti-causal. For space like coordinates, in order to preserve isotropy of space, both integrals must be combined.

The symmetric combination of $I_{+}$ and $I_{-}$  yields the Riesz integral ${_\textrm{\tiny{RZ}}}I^\alpha$:
\begin{eqnarray}
{_\textrm{\tiny{RZ}}}I^\alpha f(x) \ &=& 
{1 \over 2} (I_{+}^\alpha + I_{-}^\alpha) f(x) \\  
&=& 
\label{RZ}
\frac{1}{\Gamma(1-\alpha)}   
     \int_{-\infty}^{\infty} \!\!\!  d\xi \, |x-\xi|^{-\alpha}  f(\xi)
\end{eqnarray} 
where $||$ denotes the absolute value. These integrals are examples of a one parameter convolution integral of power law type
\begin{equation}
I^\alpha  f(x) = 
c(x)   \int_{-\infty}^\infty \!\!\!  d\xi \, w(d) f(\xi)
 \end{equation} 
with the kernel
\begin{equation}
w(d) = d^{-\alpha}
\end{equation} 
and $d$
\begin{equation}
d = |x - \xi|
\end{equation} 
is a measure of distance on $R^1$.

Erd$\acute{\text{e}}$lyi-Kober integrals are extensions of
the Riemann-Liouville  left and right fractional integrals, depending not only on the order $\alpha >0$ but also on weight $\gamma \in {\mathbb R}$ and an additional parameter $\beta>0$ as follows:
\begin{eqnarray}
\label{EK+}
I^{\alpha , \gamma }_{+;\beta} f(x) 
&=& 
\frac{x^{\beta(\gamma+1-\alpha)}}{\Gamma(1-\alpha)}
\int_0^x 
d\xi^{\beta}
(x^{\beta} - \xi^{\beta})^{-\alpha}\, \xi^{-\beta \gamma}
f(\xi)  \\
&=& 
\frac{1}{\Gamma(1-\alpha)}
\int_0^1 
d\sigma
(1-\sigma)^{-\alpha} \sigma^{-\gamma}  \, f(x
\sigma^{1/\beta}) 
\\
\label{EK-}
I^{\alpha , \gamma }_{-;\beta} f(x) 
&=& 
\frac{x^{\beta\gamma}}{\Gamma(1-\alpha)}
\int_x^{\infty} 
d\xi^{\beta}
(\xi^{\beta} -
x^{\beta})^{-\alpha}\,
|\xi|^{-\beta(\gamma+1-\alpha)} f(\xi) \\
& =& 
\frac{1}{\Gamma(1-\alpha)}
\int_1^{\infty} d\sigma
(\sigma-1)^{-\alpha}
\sigma^{-(\gamma+1-\alpha)} \, f(x
\sigma^{1/\beta}) 
\end{eqnarray}
If we take the special case $\beta=1$ and change the initial
point $0$ to $-\infty$ in (\ref{EK+}), the resulting modifications of the Erd\'elyi-Kober operators can be written as follows 
\begin{eqnarray}
\widetilde{I}^{\alpha , \gamma }_{+} f(x)  &=& 
{\frac{ x^{\gamma+1-\alpha}}{\Gamma(1-\alpha)}}
 \int_{-\infty}^x 
d\xi
(x - \xi)^{-\alpha}\,
 |\xi|^{-\gamma} f(\xi)  \\
\widetilde{I}^{\alpha , \gamma }_{-} f(x) &=& 
{\frac {x^{\gamma}}{\Gamma(1-\alpha)}}
 \int_x^{\infty} 
d\xi
(\xi - x)^{-\alpha}\,
|\xi|^{-\gamma +\alpha-1} f(\xi) 
\end{eqnarray}
which clearly shows the analogy to the fractional Liouville left and right integral definitions.

For space-like coordinates, both integrals may be combined, which yields a symmetric 
Erd\'elyi-Kober type generalized fractional integral of the form:
\begin{eqnarray}
\label{ekgen}
{_\textrm{\tiny{EK}}}I^{\alpha , \gamma }  f(x) \ &=& 
{1 \over 2} (\widetilde{I}^{\alpha , \gamma }_{+}
 + 
\widetilde{I}^{\alpha , \gamma }_{-}) f(x) \\  
&=& 
c(x)   
     \int_{-\infty}^{\infty} \!\!\!  d\xi \, |x-\xi|^{-\alpha} |\xi|^{-\gamma}  f(\xi)
\end{eqnarray} 
where $||$ denotes the absolute value. This integral is  an example for a two parameter convolution integral of power law type
\begin{equation}
I^{\alpha,\gamma}  f(x) = 
c(x)   \int_{-\infty}^{+\infty} \!\!\!  d\xi \, W(d_1,d_2) f(\xi)
 \end{equation} 
with the kernel $W$, which now consists of two factors $w_i$
\begin{equation}
\label{cas1}
W(d_1,d_2) = w_1(d_1) w_2(d_2)
\end{equation}
with
\begin{equation}
w_1(d_1) = d_1^{-\alpha} \quad  \textrm{and}  \quad w_2(d_2) = d_2^{-\gamma}
\end{equation} 
and the distances $d_i$
\begin{equation}
d_1 = |x - \xi|  \quad \textrm{and}  \quad d_2 = |\xi|
\end{equation} 
It should be mentioned, that for vanishing $\gamma$ the symmetric Erd\'elyi-Kober integral (\ref{ekgen}) smoothly reduces to the Riesz integral (\ref{RZ})   
\begin{equation}
\lim_{\gamma \rightarrow 0} {_\textrm{\tiny{EK}}}I^{\alpha , \gamma } ={_\textrm{\tiny{RZ}}}I^{\alpha}  
\end{equation} 
because the weight (\ref{cas1}) is the product of two single power law factors as a function of a measure of distance.
Values of equal weight are then determined by the equation:
\begin{equation}
d_1^{-\alpha} \, d_2^{-\gamma} =\textrm{const}  
\end{equation} 
which is from a geometric point of view when extended to $R^N$ nothing else, but the definition of Cassini ovaloids. Therefore in the next section, we will investigate the extension of the symmetric Erd\'elyi-Kober integral (\ref{ekgen}) to higher dimensions.  
\section{Symmetric Erd\'elyi-Kober integral on $R^N$}
On $R^N$ which is spanned by the coordinate set $\{x_n, n=1,...,N\}$ we define a set of $M$ foci $F_m$ at positions $\{x^F_{n m}\}$ via:
\begin{equation}
\vec{F}_m = \{x^F_{n m}\} \qquad n=1,...,N \quad m=1,...,M 
\end{equation}
The distance $r_m$ between a given focus position $\vec{F}_m$ and a given point $\vec{x}$ is the given by the Euclidean norm:
\begin{eqnarray}
r_m &=& |\vec{x} - \vec{F}_m|  \quad\quad\quad\quad\quad\quad m=1,...,M \\
&=& \sqrt{ \sum_{n=1}^{N} (x_n - x^F_{n m})^2 }
\end{eqnarray}
Introducing a corresponding set of $M$ fractional parameters $\{ \alpha_m, m=1,...,M\}$ 
we define the Cassini type weight $W$ as the product 
\begin{equation}
W(\vec{x}) = \prod_{m=1}^M (r_m)^{-\alpha_m} 
\end{equation}
and the generalized  symmetric Erd\'elyi-Kober integral on $R^N$ follows as
\begin{equation}
{_\textrm{\tiny{EK}}}I^{\{\alpha_m\}}  f(\vec{x}) = 
c  \int_{R^N} \!\!\!  d\xi^N \, W(\vec{\xi},\{\vec{F}_m(\vec{x})\}) f(\vec{\xi}),  \quad  \quad 0 < \sum_{m=1}^M \alpha_m < N
 \end{equation} 
For the one dimensional case ($N=1$) setting one focal point at $x^F_{1 1}=x$ we obtain the Riesz integral (\ref{RZ}), for 
two focal points   $x^F_{1 1}=x$ and $x^F_{1 2}=0$ we obtain up to a constant the symmetric Erd\'elyi-Kober integral (\ref{ekgen}). 

\begin{figure}
\begin{center}
\includegraphics[width=\textwidth]{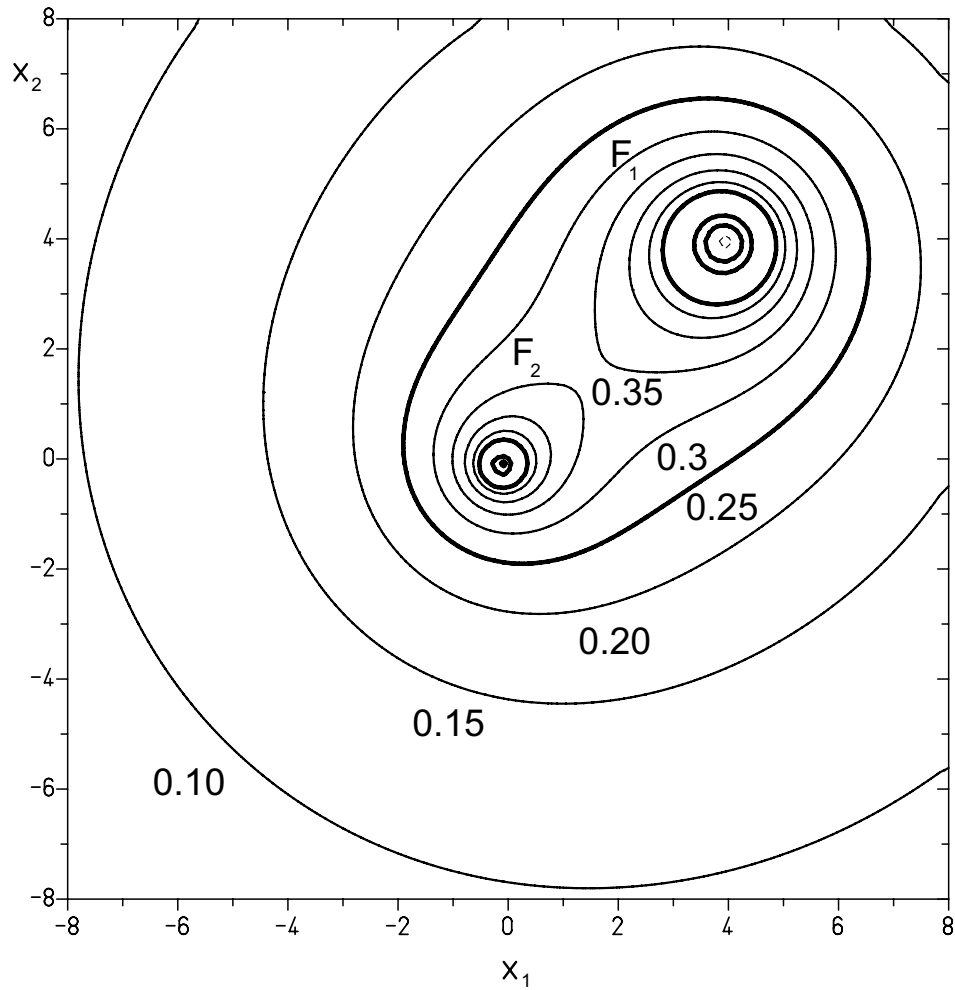}    
\caption{Contours of the weight $W$ in $R^2$ for two focal points $F_1 = \{4, 4\}$, $\alpha_1 = 0.6$ and $F_2 = \{0, 0\}$, $\alpha_2 = 0.4$. Thick lines indicate 0.25 steps.}
\label{fig1}
\end{center}
\end{figure}

In figure \ref{fig1} for the two-dimensional case we have plotted contours of the weight $W$ for two foci with two different $\alpha$.
\section{A physical interpretation and dynamic extensions}
In \cite{her11} we already mentioned, that a left handed fractional integral is causal and therefore may be used to describe the dynamics of  a particle, while the right handed fractional integral is anti-causal and may be the  appropriate tool to describe the dynamics of an anti-particle, which develops backwards in time. As a consequence,  the symmetric integral may be  used to describe particle-anti-particle pairs.

\begin{figure}
\begin{center}
\includegraphics[width=\textwidth]{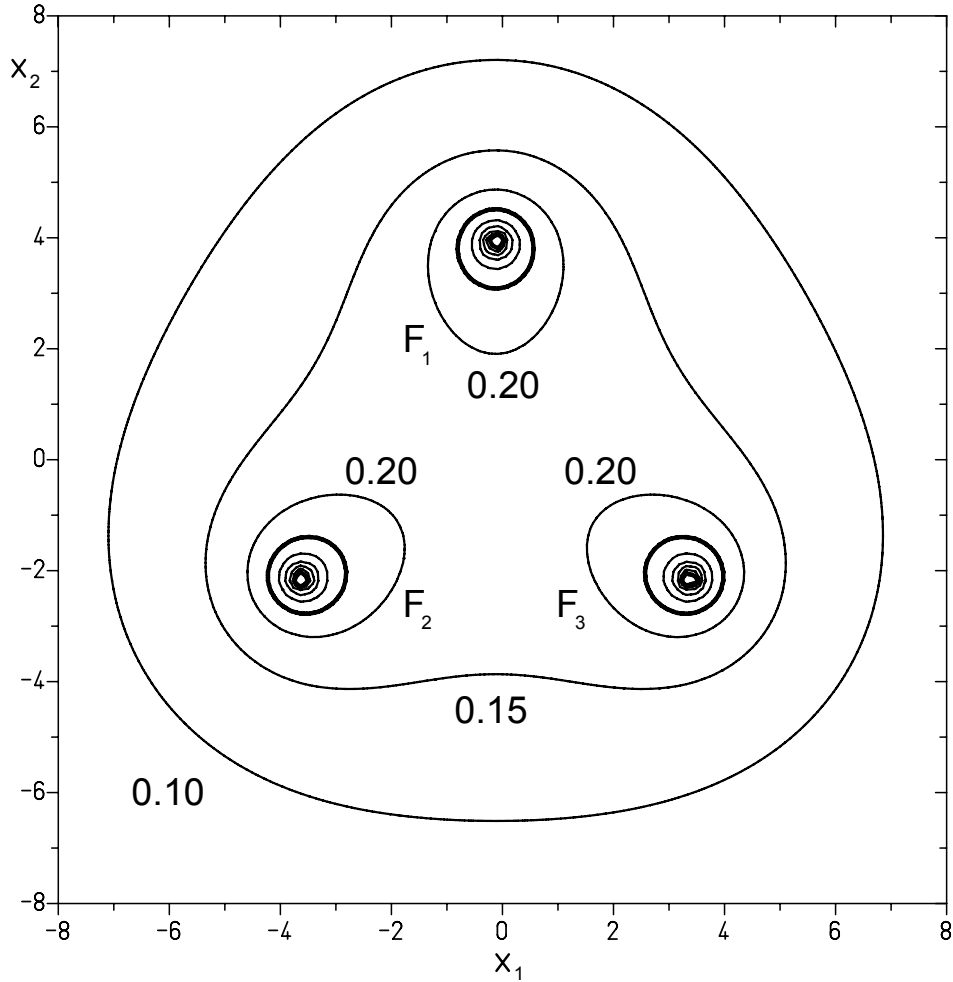}    
\caption{Contours of the weight $W$ in $R^2$ for three focal points $F_1 = \{0, 4\}$,  and $F_2$, $F_3$ rotated by $2 \pi/3$ and $4 \pi/3$ about the origin $\alpha_i = 0.4$. Thick lines indicate 0.25 steps.}
\label{fig2}
\end{center}
\end{figure}

In a similar manner the generalized symmetric Erd\'elyi-Kober type  integrals with $M$ different foci may be interpreted as operators, which describe multi-particle systems, which have a finite size. A typical example for  $M=2$ within the framework of hadron physics are mesons, which are defined as quark anti-quark systems 
\begin{equation}
m = q_1 \bar{q}_2
\end{equation}
where different charge/mass ratios are modeled using different $\alpha_1$ and $\alpha_2$ values, a method, which has already been used in nuclear physics to describe asymmetric nuclear shapes \cite{pas71}.

Excitations of such a system may then be described by a fractional differential equation of Klein-Gordon type: \begin{equation}
\big({_\textrm{\tiny{EK}}}I^{\alpha_1, \alpha_2}  \square- m^{2(\alpha_1+\alpha_2)}\big) \Psi(\vec{x}) = 0 
\end{equation}
For $M=3$ we may interpret the generalized symmetric Erd\'elyi-Kober type  as an operator suitable for a description of baryons. In figure \ref{fig2} we have sketched the weight for a symmetric configuration, which could be applied to symmetric 3-quark systems like $\Omega^- (sss)$.
Excitations of such a system may then be described by a fractional differential equation of Klein-Gordon type: \begin{equation}
\big({_\textrm{\tiny{EK}}}I^{\alpha_1, \alpha_2,\alpha_3}  \square- m^{2(\alpha_1+\alpha_2+ \alpha_3)}\big) \Psi(\vec{x}) = 0 
\end{equation}

The proposed physical interpretation also allows for an  inclusion of vibrational and rotational degrees of freedom. Until now, the presented  Erd\'elyi-Kober type operators were static. In fractional calculus, a possible time dependence of spatial operators has been discussed until now only in terms of variable order fractional parameters $\alpha \rightarrow \alpha(t) $ \cite{sam95}.  Within the framework of a geometric interpretation, a dynamic behavior may also be mediated introducing time dependent focus positions:
\begin{equation}
\vec{F}_m \rightarrow \vec{F}_m(t)  
\end{equation}
For example, the static weight shown  in figure \ref{fig2} may be extended to describe rotations 
\begin{equation}
\vec{F}_m(t)  = \hat{D}(t) \vec{F}_m(t=0) 
\end{equation}
 introducing the time-dependent rotation matrix $\hat{D} $.

Another generalization may realize the weight function $W$ not in terms of Cassini but Maxwell ovoids, which are
defined using the sum rather than the product of focal distances \cite{max46}:
\begin{equation}
\tilde{W}(\vec{x}) = \sum_{m=1}^M (r_m)^{-\alpha_m}  \qquad \qquad 0 < \alpha_m < N
\end{equation}
\section{Conclusion}
We have demonstrated that generalized classes of multi-parameter fractional integrals of power law type, which we defined as  symmetric Erd\'elyi-Kober integrals indeed may be interpreted geometrically as distortions of the rotationally symmetric kernel of Riesz fractional integrals. This interpretation allows a direct classification of higher order fractional integrals and a physical  interpretation in terms of multi-particle operators. Furthermore a new type of variable order fractional calculus in terms of space and time dependent focal position sets has been proposed.

\smallskip
\section*{Acknowledgements}
We thank A. Friedrich and V.~S. Kiryakova for useful discussions and suggestions. 



 \bigskip \smallskip

 \it

 \noindent
$^1$ GigaHedron\\
 Berliner Ring 80\\ D-63303 Dreieich
  \\[3pt]
e-mail: herrmann@gigahedron.com
\hfill Received: December 8, 2013 \\[11pt]

\end{document}